\documentclass[conference]{IEEEtran}
\IEEEoverridecommandlockouts

\usepackage{graphicx}
\usepackage{float}
\usepackage{subfigure}
\usepackage{multirow}
\usepackage{hyperref}
\usepackage{booktabs}
\usepackage{amssymb}

\usepackage{pifont} 
\usepackage{bbding} 

\usepackage{algorithm,algpseudocode} 

\AtBeginDocument{%
  }

\newcommand{\qw}[1]{\textcolor{blue}{#1}}

\usepackage{comment}
\usepackage{subfigure}

\usepackage{float}
\usepackage{fbox} 
\usepackage{fancybox}
\usepackage{framed}

\usepackage{multirow}

\usepackage{url}

\usepackage{tikz}
\usepackage{xcolor}
\definecolor{ao(english)}{rgb}{0.0, 0.5, 0.0}
\usepackage[most]{tcolorbox}

\usepackage{makecell}
\usepackage{colortbl} 

\newenvironment{packeditemize}{
	\begin{list}{$\bullet$}{
			\setlength{\labelwidth}{4pt}
			\setlength{\itemsep}{0pt}
			\setlength{\leftmargin}{\labelwidth}
			\addtolength{\leftmargin}{\labelsep}
			\setlength{\parindent}{0pt}
			\setlength{\listparindent}{\parindent}
			\setlength{\parsep}{0pt}
			\setlength{\topsep}{1pt}}}{\end{list}}

\begin{document}

\title{Deanonymizing Bitcoin Transactions via Network Traffic Analysis with Semi-supervised Learning}

\author{
\IEEEauthorblockN{Shihan Zhang$^{1}$, Bing Han$^{1}$, Chuanyong Tian$^1$, Ruisheng Shi$^{1}$, Lina Lan$^{1}$, Qin Wang$^{2}$}\\
\textit{$^1$Beijing University of Posts and Telecommunications} $|$ \textit{$^2$Independent}}

\maketitle

\begin{abstract}
Privacy protection mechanisms are a fundamental aspect of security in cryptocurrency systems, particularly in decentralized networks such as Bitcoin.  Although Bitcoin addresses are not directly associated with real-world identities, this does not fully guarantee user privacy. Various deanonymization solutions have been proposed, with network layer deanonymization attacks being especially prominent.  However, existing approaches often exhibit limitations such as low precision.

In this paper, we propose \textit{NTSSL}, a novel and efficient transaction deanonymization method that integrates network traffic analysis with semi-supervised learning. We use unsupervised learning algorithms to generate pseudo-labels to achieve comparable performance with lower costs. Then, we introduce \textit{NTSSL+}, a cross-layer collaborative analysis integrating transaction clustering results to further improve accuracy.
Experimental results demonstrate a substantial performance improvement, 1.6 times better than the existing approach using machining learning.
\end{abstract}


\section{Introduction}
Privacy protection mechanisms in cryptocurrencies are critically important. As the leading cryptocurrency, Bitcoin~\cite{nakamoto2008bitcoin} facilitates transactions without the need for trusted third parties. Bitcoin addresses used in transactions are not directly linked to real-world identities, aiming to safeguard user privacy.

However, various deanonymization solutions were proposed against Bitcoin. These attacks can be classified into two types: \textit{transaction-layer} and \textit{network-layer}.  Transaction-layer deanonymization primarily involves clustering transaction addresses that belong to the same entity, and relatively mature countermeasures, such as mixing services \cite{moser2017anonymous}\cite{coinswap}\cite{ruffing2014coinshuffle}, have been developed. In contrast, network-layer deanonymization leverages traffic transmitted over the Bitcoin network to analyze the broadcast paths of anonymous transactions, thereby linking transactions (or addresses) to real-world identifiers, such as the IP addresses of transaction originators. Notably, deanonymization attacks targeting the network layer reveal more direct and sensitive privacy information.

Classic network-layer deanonymization solutions primarily rely on analyzing transaction broadcast behaviors and timing patterns. A series of attacks~\cite{koshy2014analysis, biryukov2014deanonymisation, gao2018} attempt to identify the node that first forwards a transaction to the attacker's node, assuming that node to be the transaction's originator.

To improve user privacy, the Bitcoin community upgraded the \textit{diffusion} mechanism~\cite{bitcoin_diffusion} by introducing randomized delays before broadcasting transactions to peers, with aims to reduce the effectiveness of such deanonymization techniques.

In response, more recent deanonymization attacks targeted the diffusion mechanism. Biryukov et al.~\cite{biryukov2019deanonymization} considered the first $N$ nodes that forwarded a transaction to a listening node. This method clustered transactions that may have originated from the same node based on transaction broadcast timestamps and the IP addresses of entry nodes, but can only narrow down the range of possible IP addresses without establishing a direct link between a transaction and a specific IP. Apostolaki et al.~\cite{apostolaki2021perimeter} introduced an unsupervised learning approach to analyze Bitcoin network traffic, using an AS-level attacker to passively intercept communication between Bitcoin nodes. Unlike traditional network-layer deanonymization techniques, this approach is unaffected by changes in the transaction broadcast mechanism and can directly map transactions to the IP addresses of their originating nodes.

However, their attacks are not fully effective.

\smallskip
\noindent\textbf{Any technical challenges?} (i) If an attacker can continuously occupy all connections of a target node, it becomes trivial to analyze the node’s originating transactions. However, achieving full control over all connections (i.e., through eclipse attacks) is extremely challenging. (ii) If only a portion of the connections is occupied, can the originating transactions of the target node be inferred by analyzing its normal transaction broadcast behavior? Using a heuristic approach is very challenging, as it is difficult to derive a simple rule based on human observation.
Apostolaki et al.\cite{apostolaki2021perimeter} were the first to attempt using machine learning to \textbf{analyze the normal transaction broadcast behavior of a target node to infer its originating transactions.} 
However, the precision of this approach remains low, significantly limiting its effectiveness and its reliance on passive network-layer monitoring is undermined by Bitcoin Core v27.1’s default use of encrypted P2P transport.

\smallskip
\noindent\textbf{Contributions.} In this paper, we propose a new deanonymization method for the Bitcoin network layer based on semi-supervised learning. Our approach is compatible with the diffusion mechanism, maintains high precision in deanonymization even when only a subset of connections is occupied by ordinary attackers, and bypasses encrypted P2P transport.

\begin{figure}[!h]
  \centering
  \includegraphics[width=\linewidth]{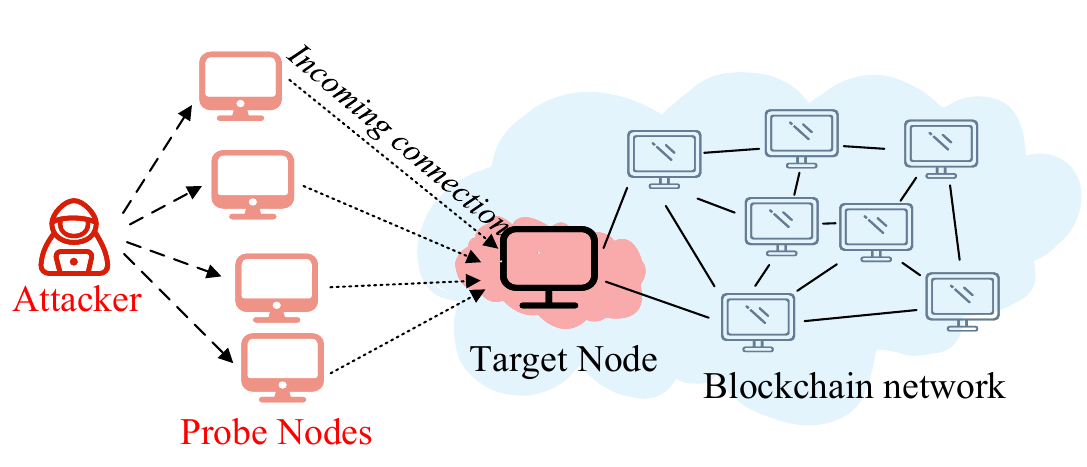}

  \caption{Attack by setting probe nodes}
  \label{fig:attack_scenario}

\end{figure}

We first collect transaction traffic data. We deploy a set of probe nodes within the Bitcoin network (cf. Fig.\ref{fig:attack_scenario}) and capture traffic by occupying the incoming connections of the target node. We analyze the traffic and observe that transactions originating from a node exhibit ``abnormal'' behavior compared to forwarded transactions. Specifically, nodes broadcast their own transactions to a larger number of peers than they do with forwarded transactions. Then, we apply a semi-supervised learning method to analyze the captured traffic, identify transactions originating from the target node, and enhance model performance by incorporating transaction clustering information from the transaction layer.

We summarize the contributions as follows.

\begin{packeditemize}
\item We propose a novel transaction deanonymization algorithm, \textit{NTSSL} (short for Network Traffic Analysis via Semi-Supervised Learning), which greatly outperforms PERIMETER~\cite{biryukov2019deanonymization}, the only known ML-based approach to date.

\item We further introduce \textit{NTSSL+}, an enhanced version that incorporates cross-layer collaborative analysis. NTSSL+ utilizes transaction clustering results from the transaction layer to refine/improve the deanonymization accuracy of NTSSL.

\item  We evaluate our deanonymization solution on the real Bitcoin network. Evaluation results show that under different percentages of connections, NTSSL achieves an F1-score ranging from 0.50 to 0.74.
NTSSL+ further improves the performance by 20\% to 40\%. 
With control over 25\% of the target node's incoming connections, NTSSL achieves about a 1.25-fold improvement compared to unsupervised learning-based deanonymization solution~\cite{apostolaki2021perimeter}, while NSSTL+ delivers a 1.6-fold improvement.

\end{packeditemize}

The rest of this paper are structured as:
Sec.\ref{sec-bkg} provides background. Sec.\ref{sec-method} describes our traffic capture method and transaction deanonymization procedure. Sec.\ref{sec-evalu} presents the experimental evaluation. Sec.\ref{sec-discussion}  discusses our findings. Sec.\ref{sec-fix} proposes our countermeasures for the identified threats. Sec.\ref{sec-rw} reviews related work.  Sec.\ref{sec-conclu} concludes this paper.

\section{Technical Warm-ups}
\label{sec-bkg}

\subsection{Bitcoin P2P Network}

\subsubsection{\underline{Bitcoin Network}} The Bitcoin network consists of nodes running Bitcoin clients that operate on a P2P network~\cite{bitcoin_protocol}. Each node both provides services to the network and utilizes services from other nodes~\cite{donet2014bitcoin}. These nodes are typically categorized into Full nodes and SPV (Simplified Payment Verification~\cite{chatzigiannis2022sok}) nodes based on the completeness of blockchain data they store and their role in transaction verification \cite{antonopoulos2023mastering}.

\begin{packeditemize} 
\item \textbf{Full nodes}. These nodes maintain a complete copy of the ledger and independently verify all blocks and transactions. They also provide additional functions, such as wallet services. Full nodes can be categorized into two types based on whether they allow incoming connections: 1) nodes that accept incoming connections and have public IP addresses, and 2) nodes that do not accept incoming connections, including those located behind NAT. Bitcoin Core is the most widely used full node implementation, accounting for over 95\% of the network’s full nodes.
\item \textbf{SPV nodes}. These nodes store only a subset of the blockchain, use Simplified Payment Verification to validate transactions. Most user wallets, particularly those on resource-limited devices, operate as SPV nodes, also known as \textit{light wallets} (Appendix~\ref{sec-lightwallet}).
\item \textbf{The relationship between Full and SPV nodes}. A transaction created by a light wallet must be validated and broadcast to other full nodes by the full node it is connected to.
\end{packeditemize}

\subsubsection{\underline{Messages}.} We present messages tracked.
\begin{packeditemize} 
\item \textbf{Inv}. A node announces transaction or block information it possesses to its neighboring nodes through an $\mathsf{Inv}$ message.
\item \textbf{Getdata}. The $\mathsf{Getdata}$ message is used to respond to an $\mathsf{Inv}$ message by requesting the object information it contains.  This message is sent after receiving the $\mathsf{Inv}$ packet and filtering out the objects the node already knows.

\item \textbf{Tx}. The $\mathsf{Tx}$ message contains the details of a Bitcoin transaction and is used to respond to a $\mathsf{Getdata}$ message from another node.

\end{packeditemize}

\subsubsection{\underline{Node Communication}}
When a new node joins Bitcoin's network, it performs network discovery and synchronizes block information by establishing connections with peers. Nodes with more blocks use the $\mathsf{Inv}$ message to propagate hash values of the available blocks. Nodes lacking these blocks can request the full block information through $\mathsf{Getdata}$ messages. 

When a node creates a transaction, it broadcasts the transaction over the P2P network. Peer nodes forward and verify it, eventually adding it to the ledger. Upon creating or receiving a new transaction, the node includes the transaction hash in an $\mathsf{Inv}$ message and sends it to its peers. When a peer first receives the $\mathsf{Inv}$ message, it responds with a $\mathsf{Getdata}$ message containing the transaction hash to request the full transaction. The original node then replies with a $\mathsf{Tx}$ message, delivering the transaction details. 

\begin{figure*}[t]
  \centering
  \includegraphics[width=\textwidth]{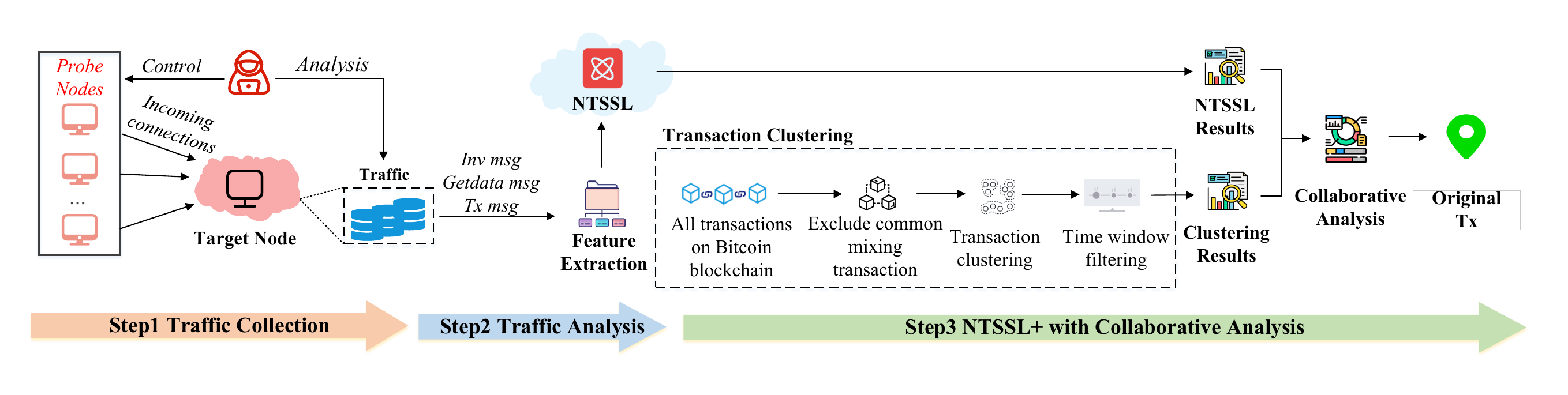}
  \caption{Attack process}
  \label{fig:attackflow}
\end{figure*}

\subsection{Additional Technical Background}

\subsubsection{Anomaly Detection}

Anomaly detection is the process of identifying objects that are different from the expected behavior pattern from a set of objects (a.k.a., anomalies or outliers). The technique is widely used in many fields~\cite{chandola2009anomaly,chatterjee2022iot} (e.g., credit card fraud detection, network security intrusion detection, industrial product fault diagnosis). Related techniques have three categories.

\begin{packeditemize}
\item \textbf{Unsupervised learning} is a type of machine learning that identifies patterns in data without labeled examples. Several popular algorithms used for unsupervised anomaly detection include Isolation Forest \cite{liu2008isolation}, Local Outlier Factor \cite{breunig2000lof}, One-class SVM \cite{heller2003one}, and Auto Encoder~\cite{jordan2018}.

\item  \textbf{Supervised learning} is a method where the model is trained on labeled data, learning to predict outcomes based on input-output pairs. Common algorithms for supervised learning include support vector machines \cite{hearst1998support}, logistic regression \cite{wright1995logistic}, and decision trees~\cite{quinlan1986induction}.

\item \textbf{Semi-supervised learning}~\cite{van2020survey,zhao2018xgbod} combines elements of both supervised and unsupervised learning. It leverages a large amount of unlabeled data while also using a smaller set of labeled data for pattern recognition. 

\end{packeditemize}

\subsubsection{Light Wallets}\label{sec-lightwallet}

Light wallets, also known as lightweight wallets, are a type of cryptocurrency wallet that does not require downloading the entire blockchain ledgers~\cite{cryptocurrency_wallet,houy2023security,chatzigiannis2022sok}. Instead, they only download block headers and rely on full nodes in the network to verify the validity of transactions. This design minimizes resource usage, making them suitable for devices with limited storage space, such as smartphones and laptops. However, these wallets differ in how they communicate with the Bitcoin network. 
Most light wallets\cite{brd_mobile,mycelium_wallet_android,samourai_wallet_tor,consenlabs_token_core,electrumx_architecture} connect directly to full nodes and their communication is typically unencrypted, posing potential risks of eavesdropping.
In contrast, light wallets like Electrum \cite{electrum_github} employ a dedicated server architecture, communicating with full nodes indirectly through backend servers. Electrum secures its communications using SSL/TLS protocols \cite{electrumx_rpc}, ensuring the safe transmission of data between the client and server. This design enhances the privacy and security of communication.

\begin{figure}[b]
  \centering
  \includegraphics[width=\linewidth]{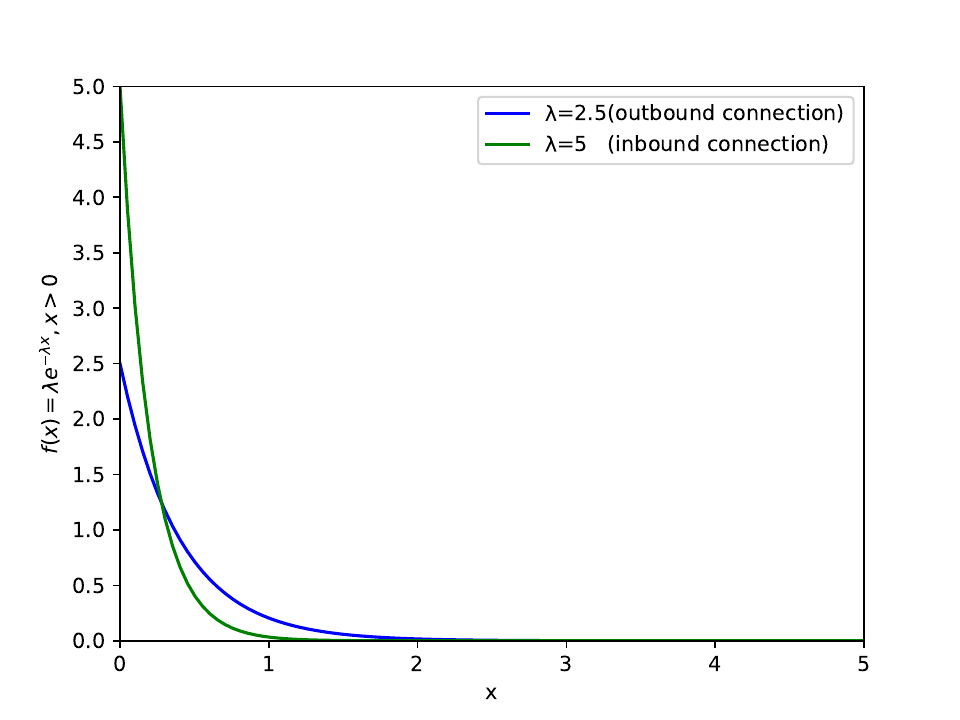}
  \caption{Comparison of random delays between outbound vs. inbound connections in the Diffusion transaction propagation mechanism}
  \label{fig:diffusion}
\end{figure}

\subsubsection{Diffusion Mechanism}\label{sec-diffusion}

In Bitcoin's diffusion transaction propagation, each node sends INV messages to its neighbors with random delays following an exponential distribution with parameter $\lambda$ (Fig.\ref{fig:diffusion}). Upon detailed analysis of Bitcoin client implementations, we found that for outbound connections, $\lambda_{\text{out}} = 2.5$ seconds; for inbound connections, $\lambda_{\text{in}} = 5$ seconds. Furthermore, when a neighboring node receives an $\mathsf{Inv}$ message, it waits for 2 seconds before sending a $\mathsf{Getdata}$ message to request the full transaction if the connection is inbound. For outbound connections, however, the $\mathsf{Getdata}$ message is sent immediately.
\section{Our Approach}
\label{sec-method}
We introduce our approach. We first present the threat model we explain the detailed attacking steps (Fig.\ref{fig:attackflow}).

\smallskip
\noindent\textbf{Threat model. }The attack goal is identify the originating transactions of the target node. The target node is a full node that accepts connections from other bitcoin nodes with a public IP address. The attacker only needs 114 nodes to establish as many incoming connections to the target node as possible.

\smallskip
\noindent\textbf{Attack process.} We have three steps. In Step-\ding{172}, we collect traffic flows from the Bitcoin network by deploying probe nodes to monitor network activity. In Step-\ding{173}, we perform traffic analysis by extracting features from the collected data and training the NTSSL model. This model classifies transactions as either node-originated or node-forwarded. In Step-\ding{174}, we by conducting cross-layer collaborative analysis by clustering transactions (i.e., NTSSL+). These clustering results are then used to refine the NTSSL classification through a collaborative analysis, enhancing the accuracy of the classification. We present details as below.

\subsection{Probe-based Traffic Collection, Step-\ding{172}}

We use a more practical end-to-end proactive traffic collection method (see Fig.\ref{fig:attack_scenario}). In this approach, the attacker uses a set of probe nodes to actively connect to the target node, occupying its incoming connections and capturing the network layer communication traffic between the target node and the probe nodes.  Since the number of incoming connections for a Bitcoin node is limited to 114 \cite{bitcoin2024reduce}, an attacker would theoretically need to deploy only 114 probe nodes to occupy all the incoming connections. Note that occupying all 114 incoming connections is not strictly required for the attacker to successfully carry out this attack. Instead, the effectiveness of the attack generally improves as the attacker controls an increasing proportion of incoming connections. Compared to the Autonomous System (AS)-based, fully passive traffic collection method proposed by Apostolaki et al. \cite{apostolaki2021perimeter}, our approach has a lower barrier to traffic collection, achievable even for ordinary attackers.

We deploy multiple fully controllable probe nodes within the Bitcoin network, acting as peers to the target node. These probe nodes actively establish connections with the target node, allowing for normal communication. As a result, they can continuously collect network layer information related to the target node, such as node version, port number, block, and transaction data. When the target node broadcasts a transaction, the probe nodes receive and record the broadcasted transaction information. The attacker's task is to identify the transactions originating from the target node within the list of transactions received by the probe nodes.

\subsection{NTSSL with Traffic Analysis, Step-\ding{173}}
We focus our analysis on three types of Bitcoin messages observed in the captured traffic: $\mathsf{Inv}$, $\mathsf{Getdata}$, and $\mathsf{Tx}$. From these messages, we extract the transaction hash and establish a mapping between the hash and the number of each message type, represented as [Hash, Num of Inv, Num of Getdata]. The transaction is then placed into an anonymity set for detection.

\subsubsection{\underline{Feature Extraction and Selection}}\label{subsec-featureselec}
We select the following features and explain the rationale. 

\begin{packeditemize}
    \item Feature 1: The total number of  $\mathsf{Inv}$ messages sent by the target node to its peers for a given transaction. Reason for selection: The target node broadcasts its originating transactions to more peers compared to forwarded transactions.

    \item Feature 2: The total number of  $\mathsf{Getdata}$ messages received by the target node from its peers for a given transaction. Reason for selection: A peer sends a  $\mathsf{Getdata}$ message only if it has not previously received the transaction. Therefore, more peers are likely to request a transaction originating from the target node, as they are less likely to have received  $\mathsf{Inv}$ messages about this transaction from other nodes.

    \item Feature 3: The ratio of the total number of  $\mathsf{Getdata}$ messages received by the target node from its peers to the total number of  $\mathsf{Inv}$ messages sent for a given transaction.Reason for selection: Due to the diffusion mechanism, the target node experiences a random delay when broadcasting a transaction to its peers. As a result, some peers may request the transaction from other nodes. In such cases, the target node may broadcast the transaction (i.e., send an  $\mathsf{Inv}$ message) but receive few or no  $\mathsf{Getdata}$ messages in return. Transactions originating from the target node are expected to have a higher request-to-broadcast ratio compared to those forwarded by the target node.

    \item Feature 4: The sum of the number of  $\mathsf{Inv}$ messages sent and  $\mathsf{Getdata}$ messages received by the target node for a given transaction.Reason for selection: This feature is primarily used to address the following scenario: the target node broadcasts two transactions, one of which is an originating transaction (Txa) and the other a forwarded transaction (Txb). Before introducing this feature, let’s assume the feature vectors of the two transactions are: Txa: [18, 6, 1/3] and Txb: [9, 3, 1/3]. From features 1 and 2, it is evident that there is a significant difference between originating and forwarded transactions. However, in this specific case, feature 3 may reduce the distinction between abnormal and normal data, thereby lowering the performance of the anomaly detection algorithm. Therefore, we propose this feature to mitigate that effect and maintain the distinction.
    
\end{packeditemize}

In addition, we augment original feature vectors using an unsupervised representation learning algorithm to further enhance the performance of our anomaly detection algorithm~\cite{chen2016xgboost}.

\smallskip
\noindent\textbf{Why not include transaction forwarding time as part of the feature vector?} Due to the diffusion mechanism in the Bitcoin network, the time of transaction broadcasts is already randomized. Simply relying on broadcast time does not provide useful information for identifying the originating transaction. Therefore, the feature of transaction broadcast time is not included. Additionally, our feature values are constructed based on the content of the traffic, whereas transaction broadcast time is an attribute independent of traffic content. Investigating how to leverage transaction forwarding time can be explored in future work.

\subsubsection{\underline{NTSSL}}
\begin{figure}[b]
  \centering
  \includegraphics[width=\linewidth]{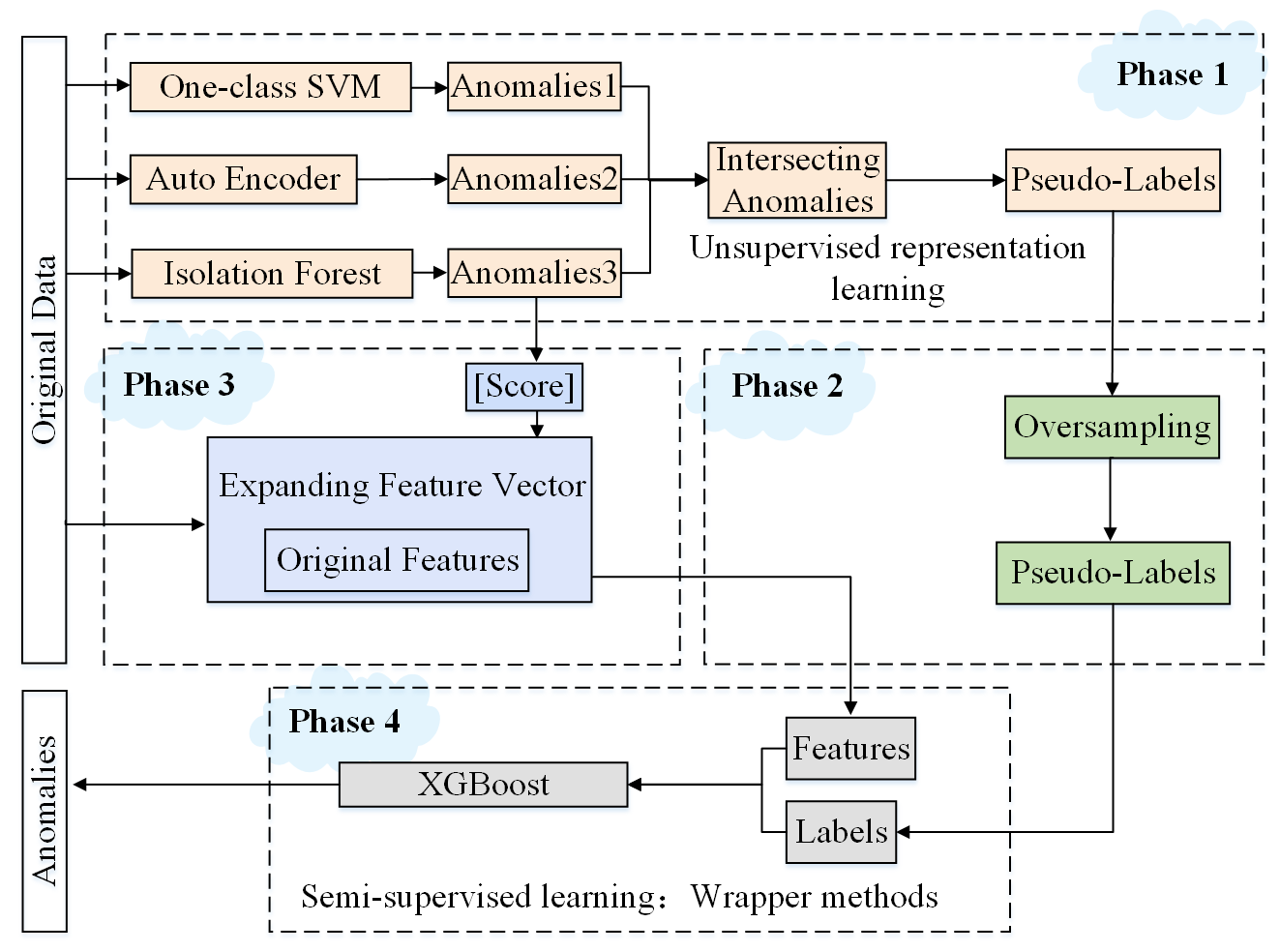}
  \caption{How our NTSSL works.}
  \label{fig:NTSSL}
\end{figure}

We propose a transaction deanonymization method based on semi-supervised learning, utilizing an "unsupervised + supervised" architecture.  In the unsupervised phase, we employ a representation learning method to perform oversampling on the training set and assign pseudo-labels. In the supervised phase, the XGBoost algorithm is used to identify transactions originating from the target node within its broadcast transactions. NTSSL has four main phases (cf. Fig.\ref{fig:NTSSL} and Algorithm~\ref{alg-ntssl}).

\smallskip
\noindent{\bfseries Phase 1: Labeling suspected anomalies.} 
We apply three unsupervised learning algorithms — Isolation Forest, Auto Encoder, and One-Class SVM — to the dataset.  Each algorithm identifies transactions it considers ``abnormal'' based on a set threshold. To improve result reliability, we take the intersection of these datasets and label the intersecting data as positive samples (i.e., anomalies). Since transactions originating from nodes are considered "abnormal" compared to forwarded transactions, they are more likely to be detected by anomaly detection algorithms. When all three algorithms identify the same transaction as originating, we can be more confident in its classification as an originating transaction.

\smallskip
\noindent{\bfseries Phase 2: Generating the final pseudo-labels.} 
The pseudo-labels assigned in the previous step are treated as known labels for the data. To address the class imbalance in the training set, we apply oversampling and use unsupervised representation learning to calculate the anomaly score for each transaction. To minimize potential false labeling during oversampling, transactions that meet all of the following conditions are labeled as positive samples (`1'), while the rest are labeled as negative samples (`0'):

\begin{packeditemize}
    
\item The anomaly score is not lower than the lowest anomaly score of the known positive samples from {\bfseries Phase 1}.

\item Send\_inv\_n is not lower than the lowest Send\_inv\_n of the known positive samples from {\bfseries Phase 1}.

\item Recv\_getdata\_n is not lower than the lowest Recv\_getdata\_n of known positive samples from {\bfseries Phase 1}.

\end{packeditemize}

According to the feature selection strategy (cf. Sec.\ref{subsec-featureselec}), the higher the Send\_inv\_n value of a sample (i.e., the more  $\mathsf{Inv}$ messages the target node sends regarding the transaction), the more likely it is that the transaction originated from the node. Similarly, the higher the Recv\_getdata\_n value of a sample (i.e., the more  $\mathsf{Getdata}$ messages the target node receives about the transaction), the more likely it is that the transaction originated from the node. Additionally, the higher the anomaly score of a sample in unsupervised learning, the more "abnormal" it is considered, and thus, the more likely it is a node-initiated transaction (i.e., a positive sample).

To summarize, the initial pseudo-label sample set obtained in the first step contains positive samples with high confidence. If additional positive samples exist outside this set (e.g., sample A), then the anomaly score, Send\_inv\_n value, and Recv\_getdata\_n value of sample A should be at least higher than the corresponding lowest values in the initial pseudo-label set. This indicates that sample A exhibits at least the same level of abnormality in its features as the samples in the initial set, further suggesting that sample A is very likely to be a positive sample.

\smallskip
\noindent{\bfseries Phase 3: Expanding the feature vector.} 
We incorporate the anomaly score obtained from the unsupervised learning algorithm as a new feature and add it to the original feature vector. Specifically, we use the anomaly score from the Isolation Forest algorithm as the additional feature, resulting in the final feature vector: [Send\_inv\_n, Recv\_getdata\_n, Recv\_getdata\_n/Send\_inv\_n, Recv\_getdata\_n+Send\_inv\_n, Score].

\smallskip
\noindent{\bfseries Phase 4: Model training and prediction.} 
We use XGBoost as the final supervised learning algorithm to train the model and generate prediction results.

\begin{algorithm}[t]
\caption{NTSSL}\label{alg-ntssl}

\renewcommand{\arraystretch}{1.0}
\scriptsize
\setlength{\tabcolsep}{2pt}
\begin{flushleft}
\quad \textbf{Input:}\textbf{$\mathcal{X}$:}Trainset raw data,${\forall x \in \mathcal{X}}$,$x \leftarrow [Inv_{Num},Getdata_{Num}]$,\\
\qquad \qquad \textbf{$X$:}Testset raw data,${\forall x \in X}$,$x \leftarrow [Inv_{Num},Getdata_{Num}]$.\\
\quad \textbf{Output:}\textbf{$\mathcal{Y}$:}Prediction of originating transactions in the test set.
\end{flushleft}

\vspace{0.1cm}

\begin{algorithmic}[1]

\State ${\forall x \in \mathcal{X}}$,$x \leftarrow [Inv_{Num},Getdata_{Num},Ratio,Sum]$
\State ${\forall x \in X}$,$x \leftarrow [Inv_{Num},Getdata_{Num},Ratio,Sum]$
\State $\mathcal{A}$ = IForest.predict($\mathcal{X}$)
\State $\mathcal{B}$ = Auto-Encoder.predict($\mathcal{X}$)
\State $\mathcal{C}$ = One-Class\_SVM.predict($\mathcal{X}$)
\State $\mathcal{P}$ = $\mathcal{A} \cap \mathcal{B} \cap \mathcal{C}$
 \Comment{\textcolor{teal}{$\mathcal{P}$ is the initial positive label set; }\textcolor{teal}{$X_p$ is identified as positive sample.}}

\noindent\makebox[\linewidth]{\rule{0.99\linewidth}{0.3pt}}

\State $\mathcal{X}_{with\_score}$ = IForest($\mathcal{X}$)
\Statex \Comment{$\mathcal{X}_{with\_score}$:[$Inv_{Num},Getdata_{Num},Ratio,Sum,Score$]}
\State $X_{with\_score}$ = IForest($X$)
\Statex \Comment{$X_{with\_score}$:[$Inv_{Num},Getdata_{Num},Ratio,Sum,Score$]}
\For{$x$ in $\mathcal{X}_{with\_score}$}
{
    \If{
    $x.Score>=min(X_{p\_Score})$ and $x.Inv_{Num}>=min(X_{p\_Inv\_Num})$ 
    \Statex \qquad \quad and $x.Getdata_{Num}>=min(X_{p\_Getdata\_Num})$ 
    }
        \State $x \rightarrow \mathcal{P}$
      \Comment{\textcolor{teal}{Set $x$ that meets the condition as a positive sample '1'.}}
        \Else
        \State $x \rightarrow \mathcal{Q}$
       \Comment{\textcolor{teal}{The remaining samples are marked with '0'.}}
    \EndIf
}
\EndFor
\State $\mathcal{L} = \mathcal{P} + \mathcal{Q}$
\State model = XGBoost($\mathcal{X}_{with\_score},\mathcal{L}$)
\State $\mathcal{Y}$= model.predict($X_{with\_score}$)

\end{algorithmic}
\label{alg:low_fee_attack}
\end{algorithm}

\subsection{NTSSL+ with Collaborative Analysis, Step-\ding{174}}

\subsubsection{\underline{Transaction Clustering}}
\label{subsec-txCluster}
Relying solely on traffic captured and analyzed at the Bitcoin network layer may not yield optimal results for deanonymization. To enhance effectiveness, we incorporate transaction-layer clustering techniques. Traditional clustering\footnote{Note that transaction clustering here refers specifically to the multi-input heuristic commonly used in cryptocurrency anonymization research (see Heuristic I—Multi-input Transactions in~\cite{androulaki2013evaluating}).} methods at the transaction layer mainly focus on grouping addresses, with the most common approach being the multi-input method, which clusters addresses belonging to the same wallet. Our method, however, shifts the focus to clustering transactions—identifying transactions entering the network from the same full node. Although the emphasis differs, the core technique remains grounded in multi-inputs.

First, to mitigate the impact of mixing transactions on clustering accuracy, we apply a filtering process to exclude common mixing transactions, partially inspired by~\cite{moser2017anonymous}. Next, we apply the multi-input method to cluster all transactions on the Bitcoin chain, obtaining preliminary clusters of transactions associated with different wallets. Since a transaction initiated by a wallet is typically first broadcast to the Bitcoin network through the full node connected to that wallet, we perform the final clustering process based on time windows (\textit{left} of Fig.\ref{fig:tx_clustering}). For example, according to the multi-input method, Tx1, Tx2, and Tx3 belong to the same wallet. If Tx2 and Tx3 are generated by the same wallet and occur within a short time frame, they are assumed to have entered the network through the same full node (i.e., both transactions are classified as originating from the same node, such as Node B).


The reason for first clustering all on-chain data before applying filtering is twofold: (i) If Tx1, Tx2, and Tx3 are not initially clustered together, Tx2 and Tx3 do not share the same input address, making the multi-input method inapplicable. (ii) If there is a significant time gap between Tx1 and Tx2/Tx3, it is highly probable that Tx1 propagated through a different full node than Tx2 and Tx3.

\subsubsection{\underline{Collaborative Analysis}}
\begin{figure}[!h]
  \centering
  \includegraphics[width=\linewidth]{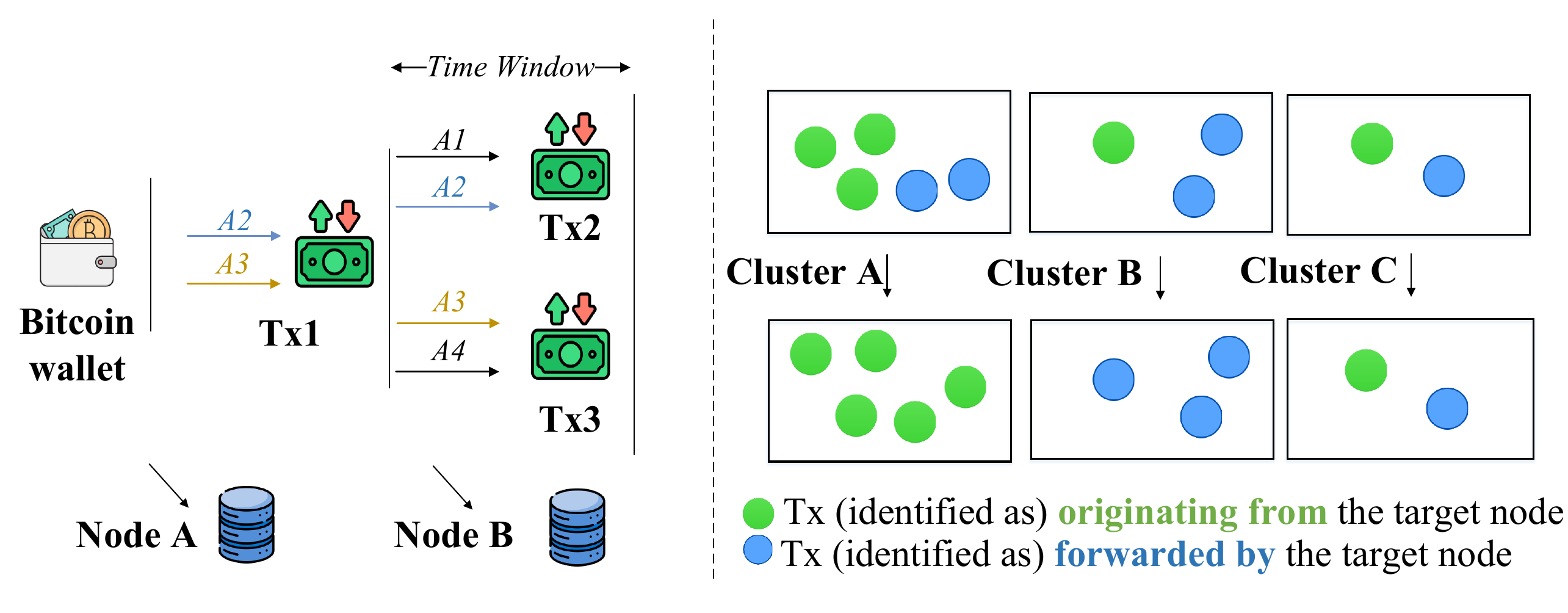}
  \caption{Collaborative analysis}
  \label{fig:tx_clustering}
\end{figure}

After executing NTSSL, we obtain initial results: \textit{a set of transactions identified as originating from the target node and a set of forwarded transactions}. To refine these results, we leverage transaction clustering to identify potential false positives and false negatives in NTSSL's output. For transactions classified by NTSSL as originating from the target node, we analyze the cluster they belong to. If the majority of transactions in the cluster are labeled as node-originated, any transactions marked as forwarded are considered algorithmic false negatives and are reclassified as node-originated. Conversely, if the majority of transactions in the cluster are labeled as forwarded, any transactions initially identified as node-originated are treated as false positives and are corrected to node-forwarded. This improves recall and reduces the false positive rate (see process in \textit{right} of Fig.\ref{fig:tx_clustering}).

\section{EVALUATION}
\label{sec-evalu}

In this section, we evaluate our transaction deanonymization approach through experiments using controllable nodes on the Bitcoin mainnet and testnet. PERIMETER~\cite{apostolaki2021perimeter} is used as the baseline (the  most relevant and only available study to ours) for comparison.

\smallskip
\noindent\textbf{Evaluation process.}
We capture traffic from our Bitcoin nodes and extract relevant features (Sec.\ref{subsec-dataover}), which serve as input for the PERIMETER, NTSSL, and NTSSL+ algorithms. The performance of these algorithms is assessed using four key metrics (Sec.\ref{subsec-metrics}): recall, false positive rate, precision, and F1-score. Additionally, we evaluate the deanonymization capability of attackers at varying interception levels (Sec.\ref{subsec-evaresult}), specifically at 25\%, 50\%, 75\%, and 100\% of the target node's connections. To realistically simulate these scenarios, we used Wireshark to capture traffic from all active connections of our self-deployed target node. We then randomly selected subsets (e.g., 25\%) of incoming connections, allowing a comprehensive assessment of threats to transaction anonymity based on the proportion of connections an attacker can control. Furthermore, to demonstrate the limitations of applying a supervised learning model trained on a single node in multi-node scenarios—and thus motivate our semi-supervised approach—we apply the XGBoost model directly on the selected feature set (Sec.\ref{subsec-featureselec}) in multi-node deployments on both Testnet and Mainnet (Sec.\ref{sec:multi_node_experiments}), omitting the unsupervised learning stage and revealing its poor cross-node generalization.



\smallskip
\noindent\textbf{Ethical considerations.}
For ethical reasons, all experiments conducted in this work strictly involve nodes we deployed ourselves.  The probe nodes act as semi-honest full Bitcoin nodes, adhering strictly to protocol specifications without introducing unwanted traffic into the Bitcoin network, thus protecting network integrity and participant privacy.

\subsection{Experimental Configurations}
We conduct experiments in two settings: (i) a single-node setup on the Bitcoin mainnet to evaluate performance of NTSSL and NTSSL+ under controlled conditions, and (ii) multi-node deployments across both the Bitcoin testnet and mainnet to demonstrate the limitations of relying solely on supervised learning methods for originating transaction identification in heterogeneous environments.

\medskip
\noindent\textbf{Single-node Setup on Mainnet.}  
A Bitcoin Core full node (v0.22.0) is deployed on Tencent Cloud (Ubuntu 18.04.4, Intel® Xeon® Gold 6133 @ 2.50 GHz, 8 GB RAM, 800 GB disk, 10 Mbps). It exposes a public IP in the Asia/Shanghai timezone (average RTT = 174 ms), runs protocol version 70016, and supports NODE\_NETWORK, NODE\_WITNESS, NODE\_NETWORK\_LIMITED.  
Configuration details are summarized in Table~\ref{tab:single_config}.

\begin{table}[ht]
  \centering
  \caption{Single-node experimental configuration}
  \label{tab:single_config}
  \begin{tabular}{l|c}
    \toprule
    \textbf{Component}       & \textbf{Specification}                \\
    \midrule
    CPU / RAM                & Xeon Gold 6133 @ 2.50 GHz / 8 GB      \\
    Disk / Bandwidth         & 800 GB SSD / 10 Mbps                 \\
    Bitcoin Core version     & v0.22.0                              \\
    Protocol version         & 70016                                \\
    Location / RTT           & Asia/Shanghai / 174 ms               \\
    \bottomrule
  \end{tabular}
\end{table}

\medskip
\noindent\textbf{Multi-node Deployments.}  
We deployed three independent full nodes on both Testnet and Mainnet. Apart from the details listed in Tables~\ref{tab:multi_testnet_node_server_configurations} and \ref{tab:multi_mainnet_node_server_configurations}, all other hardware and software configurations (e.g., OS, Bitcoin Core version and protocol version) are identical to the single-node setup.

\begin{table}[!hbt]
\caption{Multi-Node Configurations on Testnet}
\label{tab:multi_testnet_node_server_configurations}

\centering
\resizebox{0.9\linewidth}{!}{
\begin{tabular}{c|c|c|c}
\toprule
\textbf{Node} & \textbf{Public IP} & \textbf{\textcolor{teal}{Testnet} Configuration} & \textbf{Bandwidth} \\
\midrule
Node A & 43.143.185.122 & 2 cores, 4 GB RAM, 1000 GB disk & 6 Mbps \\
Node B & 43.143.203.198 & 2 cores, 4 GB RAM, 1000 GB disk & 6 Mbps \\
Node C & 101.43.124.195 & 4 cores, 8 GB RAM, 800 GB disk & 10 Mbps \\
\bottomrule
\end{tabular}
}
\end{table}

\begin{table}[!hbt]
\caption{Multi-Node Configurations on Mainnet}
\label{tab:multi_mainnet_node_server_configurations}
\centering
\resizebox{0.9\linewidth}{!}{
\begin{tabular}{c|c|c|c}
\toprule
\textbf{Node} & \textbf{Public IP} & \textbf{\textcolor{teal}{Mainnet} Configuration} & \textbf{Bandwidth} \\
\midrule
Node A & 43.143.185.122 & 2 cores, 4 GB RAM, 1000 GB disk & 6 Mbps \\
Node C & 101.43.124.195 & 4 cores, 8 GB RAM, 800 GB disk & 10 Mbps \\
Node D & 118.25.12.97 & 4 cores, 8 GB RAM, 800 GB disk & 10 Mbps \\
\bottomrule
\end{tabular}
}
\end{table}

\subsection{Data Overview}\label{subsec-dataover}

\medskip
\noindent\textbf{Single-node Dataset.} 
In total, we captured $\sim$15K transactions, of which $\sim$90 originated from the target node itself. 
For each transaction, we extracted the following features: the number of  $\mathsf{Getdata}$ messages received by the node, denoted as Getdata\_Num; the number of  $\mathsf{Inv}$ messages sent by the node, denoted as Inv\_Num; the ratio of these two values; and the sum of these two values. These features were combined to generate the original feature vector \(X\):
\[
X = \begin{bmatrix}
\mathsf{Inv\_Num}, \mathsf{Getdata\_Num}, \mathsf{Ratio}, \mathsf{Sum}
\end{bmatrix}.
\]
The original feature space that contains all \(n\) transactions is:
\[
\mathsf{Feature\ Space_{original}} = 
\begin{bmatrix}
X_1 \\
X_2 \\
\vdots \\
X_n
\end{bmatrix} \in M^{n \times d}(\mathbb{R}),
\]
where each \(X_i\) (for \(i = 1, 2, \dots, n\)) corresponds to the feature vector of an individual transaction, and \(d = 4\) is the number of features.
In unsupervised learning stage, we extend each \(X_i\) by adding its anomaly score from the Isolation Forest algorithm, denoted as \(\mathsf{Score\ IF}(X_i)\). The new feature vector becomes:
\[
X_{\text{new}, i} = \begin{bmatrix}
\mathsf{Inv\_Num}, \mathsf{Getdata\_Num},  \mathsf{Ratio},  \mathsf{Sum},  \mathsf{Score\ IF}(X_i)
\end{bmatrix}.
\]
The corresponding new feature space is:
\[
\mathsf{Feature\ Space_{new}} = 
\begin{bmatrix}
X_{\text{new}, 1} \\
X_{\text{new}, 2} \\
\vdots \\
X_{\text{new}, n}
\end{bmatrix} \in M^{n \times (d + 1)}(\mathbb{R}),
\]
where each row corresponds to the extended feature vector of a transaction.
The feature space \(\mathsf{Feature\ Space_{new}}\) will be applied to our deanonymization approach based on semi-supervised learning. To ensure result reliability, we ran the unsupervised anomaly detection algorithm five times on the experimental dataset and used the average outcomes as the final performance metric. Additionally, we employed a 5-fold cross-validation method to evaluate the performance of the NTSSL algorithm and model.

\medskip
\noindent\textbf{Multi-node Datasets.}
Table~\ref{tab:testnet_node_dataset_details} summarizes per-node traffic statistics. On Testnet, each node contributes roughly 4.5K$\sim$8K transactions for training and 3.8K$\sim$6.9K for testing, with 40 originating transactions per dataset. On Mainnet, per‐capture volumes span approximately 5.8K$\sim$54K transactions, with 20$\sim$44 originating transactions per node. In these multi-node experiments, we restrict ourselves to the original 4-dimensional feature vectors \(\bigl[\mathsf{Inv\_Num},\,\mathsf{Getdata\_Num},\,\mathsf{Ratio},\,\mathsf{Sum}\bigr]\) and do not apply the Isolation Forest scoring used in the single-node semi-supervised pipeline.

\subsection{Metrics}\label{subsec-metrics}

We selected \textit{recall}, \textit{false positive rate}, \textit{precision}, and \textit{F1-score} as evaluation criteria, which are widely used in the field of machine learning to assess classification model performance \cite{ng_ml_course}. We first define \textbf{confusion matrix} as below. TP represents True Positives, FN represents False Negatives, FP represents False Positives, and TN represents True Negatives.

\begin{table}[h!]
\centering
\resizebox{0.75\linewidth}{!}{
\begin{tabular}{c|cc}
\toprule
\textbf{True Value / Predicted Value} & \textbf{Positive} & \textbf{Negative} \\ 
\cmidrule{1-3}
\textbf{Positive} &  \cellcolor{gray!15} TP &  \cellcolor{gray!15} FN \\ 
\textbf{Negative} &  \cellcolor{gray!15} FP &  \cellcolor{gray!15} TN \\ 
\bottomrule
\end{tabular}
}
\end{table}

We then give the definitions based on confusion matrix.

\begin{packeditemize}

\item \textbf{Recall} measures the proportion of actual positive samples that are correctly predicted as positive. In our case, it refers to the proportion of correctly identified transactions among all transactions originating from the node. Higher recalls indicate better.
\begin{equation*}
  \mathsf{Recall} = \frac{\mathsf{TP}}{\mathsf{TP} + \mathsf{FN}}
\end{equation*}

\item \textbf{False positive rate (FPR)} is the proportion of negative samples incorrectly predicted as positive. In our context, FPR represents the proportion of forwarded transactions wrongly identified as coming from the target node. Lower FPRs indicate better.
\begin{equation*}
  \mathsf{FPR} = \frac{\mathsf{FP}}{\mathsf{FP} + \mathsf{TN}}
\end{equation*}

\item \textbf{Precision} measures the accuracy of positive predictions. In this context, it refers to the proportion of transactions from the target node within the identified transactions. Higher indicates better. 
\begin{equation*}
  \mathsf{Precision} = \frac{\mathsf{TP}}{\mathsf{TP} + \mathsf{FP}}
\end{equation*}

\item \textbf{F1-score} combines recall and precision into a single metric, calculated as their harmonic mean. It provides the measure of the model’s performance. Higher indicate better.
\begin{equation*}
  \mathsf{F1\text{-}score} = 2 \times \frac{\mathsf{Precision} \times \mathsf{Recall}}{\mathsf{Precision} + \mathsf{Recall}}
\end{equation*}

\end{packeditemize}



\subsection{Evaluation Results (Fig.\ref{fig:evalution})}\label{subsec-evaresult}

\subsubsection{\underline{PERIMETER}.}
From the detection results, PERIMETER does not achieve notable performance, with recall and precision both around 53\%, even when the attacker intercepts 100\% of the target node’s connections. Due to the large number of negative samples in this scenario, the false positive rate remains low, even with a certain number of false positives. As a result, we do not prioritize the false positive rate as a key metric for evaluating algorithm performance. Instead, we use the F1-score, which takes both recall and precision into account. The F1-score for this algorithm reaches 0.53.

\subsubsection{\underline{NTSSL}}
When the attacker intercepts 100\% of the target node’s connections, the NTSSL algorithm identified 78 transactions as originating from the target node within the anonymous set.  Of these, 61 were actual target node-originated transactions, while the remaining 17 were transactions forwarded by the target node but incorrectly identified as originating from it.

\begin{table}[!hbt]
  \caption{Top 10 cluster sizes with their respective amount}
  \label{tab:Top_10_cluster_size}
  \resizebox{\linewidth}{!}{
  \begin{tabular}{c|c ccc ccc ccc }
    \toprule
    \multicolumn{1}{c|}{No. of Clusters}   & 1 & 1 & 1 & 1 & 1 & 1 & 1 & 1 & 1 & 4    \\
    \cmidrule{1-11}
    Cluster Size &  \cellcolor{gray!15} 97 &  \cellcolor{gray!15} 70 &  \cellcolor{gray!15} 48 &  \cellcolor{gray!15} 47 &  \cellcolor{gray!15} 39 &   \cellcolor{gray!15} 38 &  \cellcolor{gray!15} 37 &  \cellcolor{gray!15} 35 &  \cellcolor{gray!15} 31 &   \cellcolor{gray!15} 26  \\
  \bottomrule
\end{tabular}
}
\end{table}

When the attacker intercepts 100\% of the target node’s connections, the transaction deanonymization model produced by the NTSSL algorithm achieves a recall of 70.2\%, a precision of 79\%, and an F1-score of 0.74—higher than the unsupervised learning-based deanonymization method. Even for the weakest attacker, capable of intercepting only 25\% of the target node’s connections, the algorithm still achieved a recall of 56.8\%, a precision of 49.8\%, and an F1-score of 0.5. While there is still room for optimization in this semi-supervised learning-based deanonymization method, it demonstrates a significant improvement compared to PERIMETER.

\subsubsection{\underline{NTSSL+}}
We applied the transaction clustering method (cf. Sec.\ref{subsec-txCluster}) to the dataset, resulting in a total of 12,413 clusters. The largest cluster contains up to 97 transactions. Table~\ref{tab:Top_10_cluster_size} presents the top 10 cluster sizes along with their respective transaction amount.

\begin{figure*}[t]
    \centering
    \subfigure[{25\% Connections}]{\label{fig:}
        \includegraphics[width=0.23\linewidth]{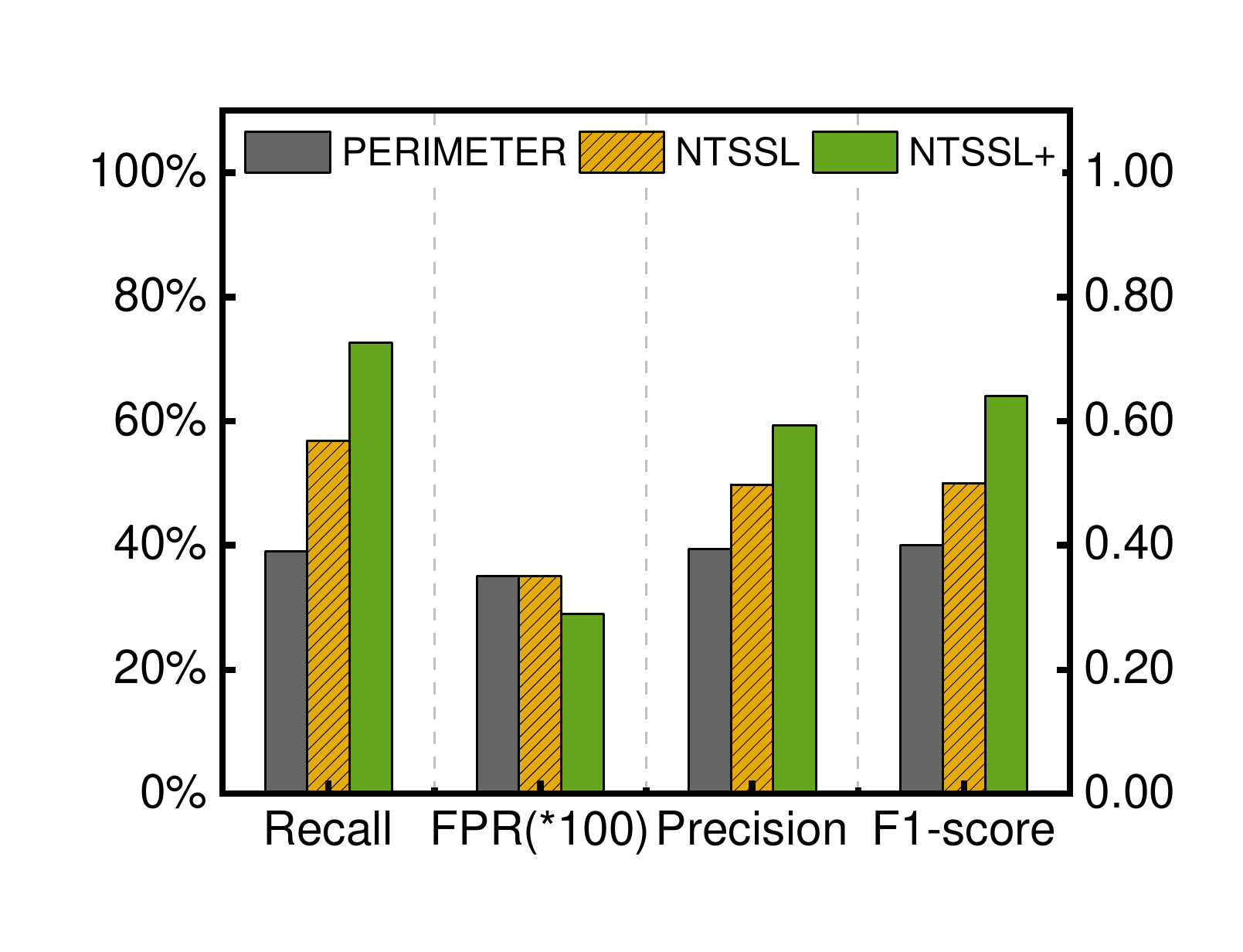}
    }
    \subfigure[{50\% Connections}]{\label{fig:}
        \includegraphics[width=0.23\linewidth]{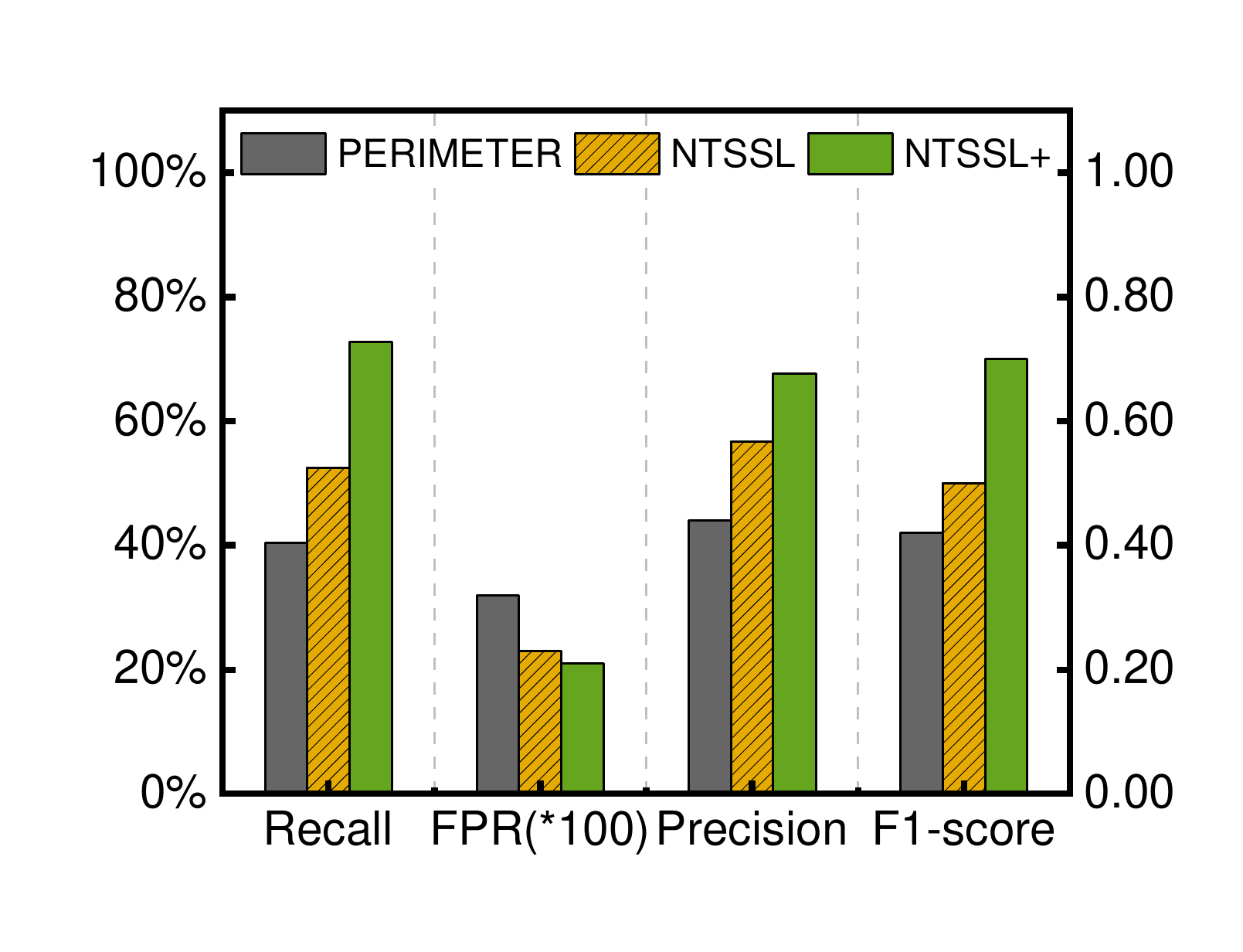}
    }  
    \subfigure[{75\% Connections}]{\label{fig:}
        \includegraphics[width=0.23\linewidth]{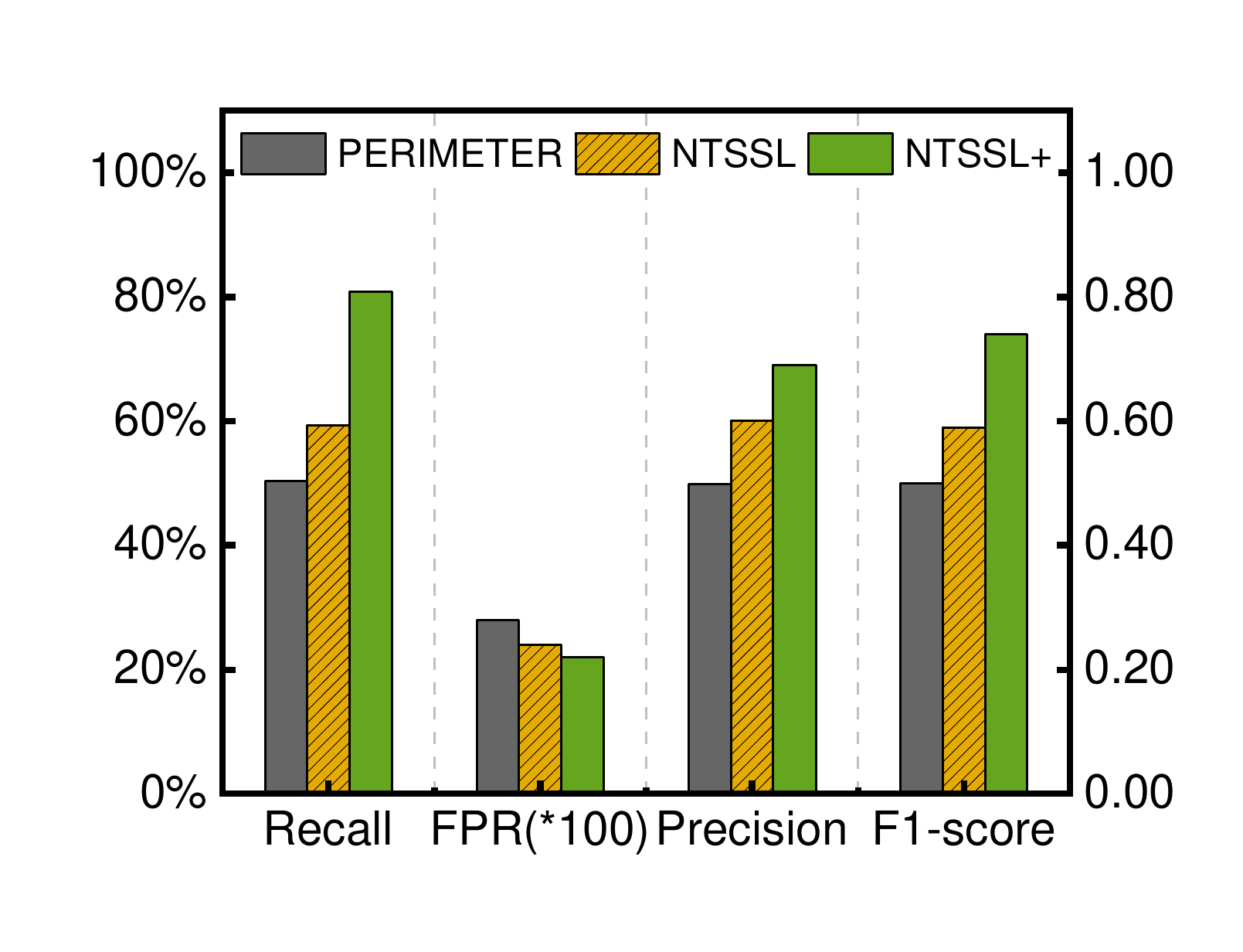}
    }
    \subfigure[{100\% Connections}]{\label{fig:}
        \includegraphics[width=0.23\linewidth]{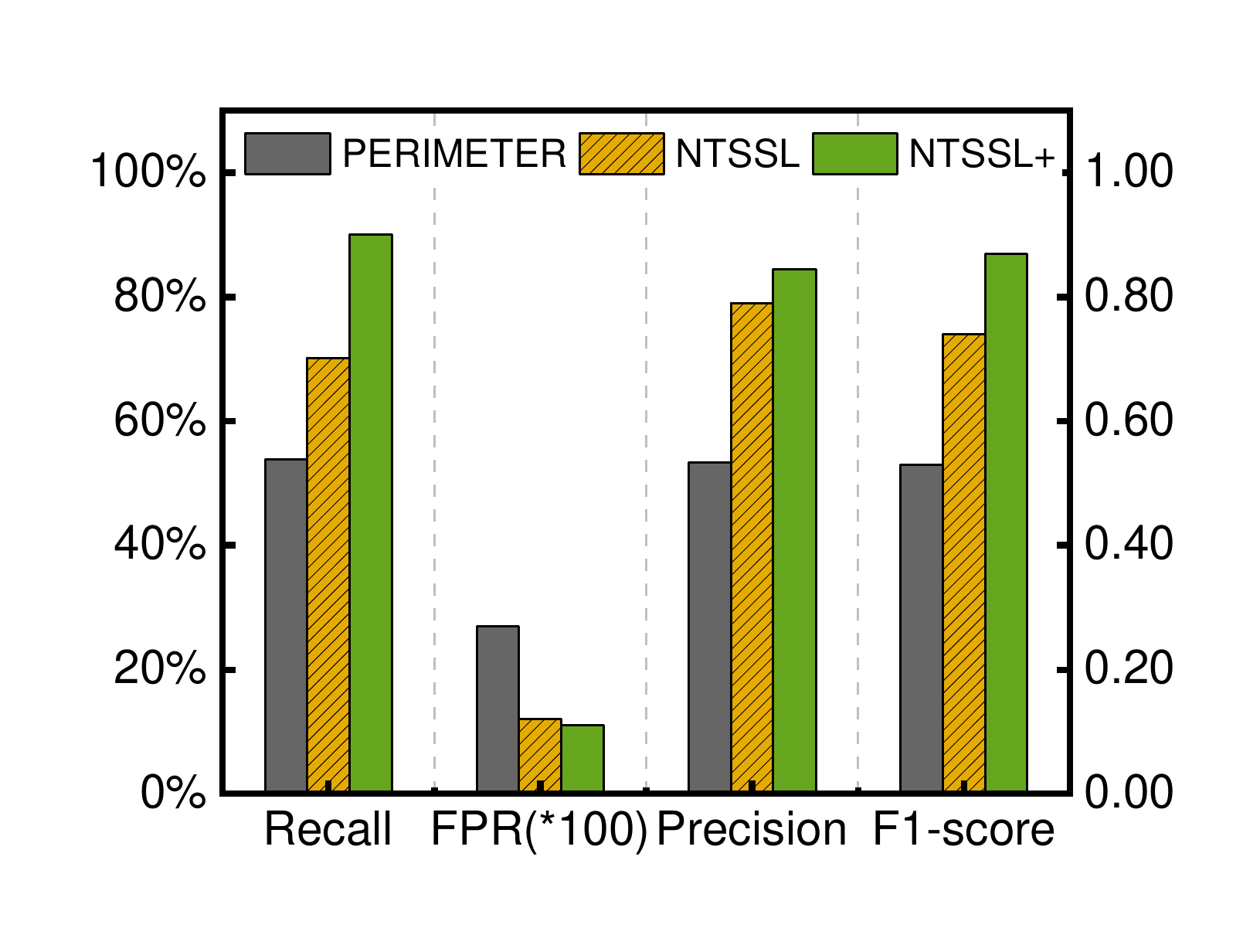}
    }
    \caption{Performance evaluation: NTSSL/NTSSL+ outperform PERIMETER across all four metrics — recall, false positive rate, precision, and F1-score — at every level of intercepted connections, with particularly outstanding performance in recall. As the number of connections increases from 25\% to 100\%, the performance of NTSSL/NTSSL+ improves steadily.
    }
    \label{fig:evalution}
\end{figure*}

Using the transaction clustering results, we identified 18 transactions originating from the target node that were not detected by the NTSSL algorithm (i.e., false negatives) and 2 node-forwarded transactions that were incorrectly classified as originating transactions (i.e., false positives). We performed similar cross-layer collaborative analysis in three other scenarios. In the scenario where the attacker intercepts 75\% of the node's connections, we found 19 false negatives and 3 false positives in the NTSSL results. When the attacker intercepts 50\% of the node's connections, there were 18 false negatives and 4 false positives. Finally, in the scenario where the attacker intercepts 25\% of the connections, we identified 16 false negatives and 9 false positives. A summary of the final performance of the cross-layer collaborative analysis is presented in Fig.\ref{fig:evalution}.


We found that incorporating transaction clustering into our deanonymization method significantly improved its performance. This improvement is primarily due to the ability to identify false positives and false negatives in the NTSSL algorithm using the transaction clustering results, which enhances both recall and precision, ultimately increasing the algorithm's F1-score. Even for the weakest attacker, capable of intercepting only 25\% of the target node’s connections, our cross-layer collaborative analysis method achieves an F1-score of 0.64. In comparison, the unsupervised learning-based deanonymization method achieves an F1-score of 0.4, while NTSSL reaches 0.5 under the same attack conditions. Overall, our cross-layer collaborative deanonymization method improves performance by 20\% $\sim$ 40\% compared to NTSSL, and far surpasses the performance of unsupervised learning-based deanonymization methods.

\subsubsection{\underline{Result Analysis}} We provide a detailed discussion on why our solution outperforms PERIMETER, highlighting the key factors contributing to its superior performance.

\smallskip
\noindent\textbf{Why PERIMETER does not perform well?}
First, the anomalies detected by the anomaly detection algorithm may represent abnormal data in different dimensions. The algorithm might flag both maximum and minimum values in the data sample as anomalies. In the context of transaction traceability, only the maximum value may correspond to the originating transaction — where the node sends more  $\mathsf{Inv}$ messages and receives more  $\mathsf{Getdata}$ messages. In contrast, the other extreme (fewer  $\mathsf{Inv}$ and  $\mathsf{Getdata}$ messages) does not represent an originating transaction. For a binary classification problem, the anomaly detection algorithm cannot distinguish between these two types of anomalies, as both receive high anomaly scores. This leads to some anomalies being incorrectly labeled as originating transactions.

Secondly, upon analyzing the data, we observed that for some nodes, the number of  $\mathsf{Getdata}$ and  $\mathsf{Inv}$ messages corresponding to the originating transaction is similar to that of the node's forwarded transactions. The anomaly detection algorithm may classify them as the same type (either both originating or both non-originating), leading to false positives or false negatives in PERIMETER.

\smallskip
\noindent\textbf{Why not use supervised learning methods?}
Through experiments (cf. Sec.~\ref{sec:multi_node_experiments}), we found that a supervised learning model trained on one node does not perform well when applied to other nodes. Upon examining the transaction data extracted from the traffic of these nodes, we identified some reasons for this issue.  In terms of feature vectors, there are noticeable differences in the transaction data across different nodes (e.g., the average number of  $\mathsf{Getdata}$ and  $\mathsf{Inv}$ messages for originating and non-originating transactions). These variations arise from differences in the number of connections and the activity levels of peers, leading to discrepancies in the number of transaction messages processed by each node over a given period. As a result, the overall transaction data from different nodes can vary significantly, indicating that they are different types of nodes. Due to these differences in transaction data, a supervised learning model trained on one node struggles to generalize to the data distributions of other nodes. In other words, the same forwarding transaction may exhibit different characteristics across different types of nodes, which explains why a supervised learning model trained on one node does not perform well on others.

Although supervised learning models are not suitable for the scenario in this paper, the work of supervised learning on transaction deanonymization is worth studying in the future.

\subsection{Limitation of Supervised Model in Multi-node Scenarios}
\label{sec:multi_node_experiments}


To illustrate the limitations of supervised learning methods in the context of originating transaction identification, we further evaluate the applicability of a supervised learning model trained on a single node to multi-node scenarios (e.g., three nodes) by conducting experiments on the Bitcoin testnet and mainnet. In these multi-node experiments, we train an XGBoost classifier directly on the original 4-dimensional feature set without any unsupervised learning component.

\subsubsection{Evaluation on the Testnet}

We conducted experiments on three different nodes in the testnet (configurations as Table~\ref{tab:multi_testnet_node_server_configurations}), with the collected data shown in \textit{upper} Table~\ref{tab:testnet_node_dataset_details} and results in \textit{upper} Table~\ref{tab:Performance_Testnet}. 

The results demonstrate that the supervised learning model trained on a single node generalizes well to other nodes within the testnet. This can be attributed to the fact that, compared to the mainnet, the testnet features fewer nodes and lower network complexity. Furthermore, the reduced number of transactions in the testnet simplifies the communication patterns for exchanging transaction information between peer nodes. Consequently, the characteristics of originating transactions in the testnet environment become more pronounced (based on the feature vectors we proposed), making them easier for the model to predict accurately.

\subsubsection{Evaluations on the Mainnet}

We conducted experiments on three different nodes in the mainnet (setup as Table~\ref{tab:multi_mainnet_node_server_configurations}). A supervised learning model trained on Node D was applied to the test datasets collected from Nodes A and C.

The results (\textit{bottom} Table~\ref{tab:Performance_Testnet}) indicate that the supervised learning model trained on a single node performs poorly when applied to other nodes in the mainnet. For example, based on the analysis of all 40,234 transactions in Node A’s dataset (\textit{bottom} Table~\ref{tab:testnet_node_dataset_details}), there are at least 690 transactions with similar characteristics to the 20 originating transactions. This significantly affects the model's ability to correctly evaluate the test data.

\begin{table}[!hbt]
\caption{Average  $\mathsf{Inv}$ and $\mathsf{Getdata}$ Counts of Nodes}
\label{tab:mainnet_node_average_counts}
\centering
\renewcommand{\arraystretch}{1.5} 
\Large 
\resizebox{\linewidth}{!}{ 
\begin{tabular}{c|cc|cc}
\toprule
\textbf{Node} & \textbf{$\mathsf{Inv}$/Tx} & \textbf{$\mathsf{Getdata}$/Tx} & \textbf{$\mathsf{Inv}$/OriginalTx} & \textbf{$\mathsf{Getdata}$/OriginalTx} \\
\midrule
 Node A & 25.258 & 1.954 & 34.3 & 11.5 \\
Node C & 12.309 & 1.072 & 10.27 & 5.13 \\
 Node D (Training Set) & 15.911 & 0.018 & 21.59 & 1.20 \\
\bottomrule
\end{tabular}
}
\end{table}

A deeper reason for the poor performance is that the nodes in the mainnet are no longer of the same type. During traffic capture, Node A consistently maintained a higher number of outgoing connections than Nodes C and D. Additionally, as shown in Table~\ref{tab:mainnet_node_average_counts}, Node A exhibited more  $\mathsf{Getdata}$ and  $\mathsf{Inv}$ messages in its transactions compared to those in the training datasets of Nodes C and D. This suggests that Node A, Node C, and Node D are not the same type of node, and a model trained on one type of data is not suitable for detecting patterns in another type.

\begin{table}[!hbt]
\caption{Node Dataset and Traffic Details}
\label{tab:testnet_node_dataset_details}

\centering
\resizebox{\linewidth}{!}{
\begin{tabular}{c| c|c|ccc|c}
\toprule
   \multicolumn{1}{c}{} &  \multicolumn{1}{c}{\textbf{Node}} &  \multicolumn{1}{c|}{\textbf{Dataset}} &  \multicolumn{1}{c}{\textbf{Total Txs}} &  \multicolumn{1}{c}{\textbf{Original Txs}} &  \multicolumn{1}{c}{\textbf{Duration}} &  \multicolumn{1}{c}{\textbf{Connections}} \\
\midrule

 \multirow{7}{*}{\rotatebox{90}{\textcolor{teal}{\textbf{Testnet}}}}  &  \multirow{2}{*}{Node A} 
     & Training Set & 8043 & 40 & 26h 14min & 10--11, all outgoing \\
   &  & Test Set     & 6913 & 40 & 26h 31min & 10--11, all outgoing \\
\cmidrule{5-7}

   &  \multirow{3}{*}{Node B} 
    &   Training Set & 4504 & 40 & 26h 22min & 10--11, all outgoing \\
  &   & Test Set     & 3799 & 40 & 24h 25min & 10--11, all outgoing \\
\cmidrule{5-7}

    &  \multirow{2}{*}{Node C} 
    & Training Set & 8031 & 40 & 26h 12min & 10--13, all outgoing \\
    & & Test Set     & 6910 & 40 & 26h 31min & 10--13, all outgoing \\
    
\midrule

\multirow{4}{*}{\rotatebox{90}{\textcolor{teal}{\textbf{Mainnet}}}} & 
Node A & Test Set & 40234 & 20 & 1h 31min & 55--75, 10 outgoing \\
& Node C & Test Set & 54382 & 30 & 1h 37min & 25--35, 10 outgoing \\
&  Node D & Training Set & 13281 & 44 & 1h 12min & 34--37, 10 outgoing \\
& Node D & Test Set    & 5831  & 21 & 30min    & 30--35, 10 outgoing \\
\bottomrule
\end{tabular}
}
\end{table}

\begin{table}[!hbt]
\centering
\caption{Performance Evaluation}
\label{tab:Performance_Testnet}

\resizebox{\linewidth}{!}{ 
\begin{tabular}{c|l|l|ccc|c}
\toprule
 \multicolumn{1}{c}{} & \multicolumn{1}{c|}{\textbf{Model}} &  \multicolumn{1}{c|}{\textbf{Testset}} & \textbf{Recall} & \textbf{FPR} & \textbf{Precision} & \textbf{F1-score} \\
\midrule
\multirow{6}{*}{\rotatebox{90}{\textcolor{teal}{\textbf{Testnet}}}} &  Model A & Testset B &   92.50\% &     0.00\% &     100.00\% &     96.10\% \\
&  Model A & Testset C &   90.00\% &     0.00\% &     100.00\% &     94.74\% \\
\cmidrule{5-7}
& Model B & Testset A &   82.50\% &     0.22\% &     68.75\% &     75.00\% \\
& Model B & Testset C &   90.00\% &     0.02\% &     97.30\% &     93.51\% \\
\cmidrule{5-7}
&  Model C & Testset A &   82.50\% &     0.00\% &     100.00\% &     90.41\% \\
&  Model C & Testset B &   92.50\% &     0.00\% &    100.00\% &     96.10\% \\
\midrule
\multirow{2}{*}{\rotatebox{90}{\textcolor{teal}{\textbf{Mn.}}}} &  Model D & Testset A &   15.00\% &     0.045\% &     14.29\% &     14.63\% \\
& Model D & Testset C &   15.00\% &     0.01\% &     30.00\% &     15.00\% \\
\bottomrule
\end{tabular}
}
\end{table}

\section{DISCUSSION}
\label{sec-discussion}

\smallskip
\noindent\textbf{How many connections can an attacker occupy?}
Experimental results show that the more connections an attacker controls, the more effective the de-anonymization process becomes. To understand the potential extent of this control, we begin with a rough estimation. The Bitcoin network has approximately 57,000 nodes, of which about 19,000 full nodes are capable of accepting incoming connections\footnote{https://bitnodes.io/}. From this, we can estimate the average number of incoming connections per node in the network as: \(\frac{57,000 \times 10} { 19,000} = 30\). This suggests that an attacker could potentially occupy a substantial portion of the connections — up to \( \frac{114 - 30}{124} = 67.74\% \) of the available slots\footnote{By default, Bitcoin Core allows up to 125 connections per node, including 10 outgoing connections and 1 feeler connection, leaving 114 available for incoming connections.}. Moreover, a clever attacker can exploit Bitcoin nodes' connection eviction mechanism to further increase control over a target node’s incoming connections \cite{saad2021syncattack,ha2023sustainability}. An attacker can even leverage the peerlist-filling attack \cite{heilman2015eclipse,tran2020stealthier} to occupy a portion of the target node’s outgoing connections, although fully taking over all outgoing connections remains challenging.


\smallskip
\noindent\textbf{How to determine the size of time window?} 
In general, users create transactions through either a lightweight wallet or a full-node wallet (e.g., a Bitcoin Core full node behind NAT). During a user session (i.e., the interaction between the  wallet and the node), the transactions within a cluster (identified using the common input method) are propagated to the network via the full node.

An attacker can determine the appropriate time window size by analyzing gateway traffic to gather statistical data on the duration of sessions between the lightweight wallet and the target full node. By examining the distribution of session durations, the attacker can then estimate the optimal size for the time window.


\smallskip
\noindent\textbf{Encrypted traffic.} Starting with the release of Bitcoin Core version v26.0 in December 2023, Bitcoin has experimentally supported the use of the version 2 (v2) P2P cryptographic transport protocol \cite{bip0324} between nodes. As of Bitcoin Core version v27.1, released on June 17, 2024, v2 P2P encryption has become the default protocol for communication between Bitcoin nodes. While this encrypted transmission protocol mitigates the deanonymization method proposed by Apostolaki M et al. \cite{apostolaki2021perimeter}, it does not affect our deanonymization approach. In our method, a probe node is used to establish a direct connection with the target node, collecting plaintext communication traffic, thus bypassing the encryption.

\smallskip
\noindent\textbf{Extreme case.} If the target node lacks originating transactions, how can the effectiveness of NTSSL model training be ensured? An attacker can manually configure the full node under their control to exclusively connect to the target node and send newly created transactions to it. This guarantees that the target node is the first to propagate the transaction to the Bitcoin network, providing the NTSSL model with originating transaction samples, thereby ensuring effective model training.

\smallskip
\noindent\textbf{The attack we propose is highly scalable.} It can be easily expanded across the entire network without requiring additional resources, as the previously deployed probe nodes can be reused. Moreover, the attacker can strategically target key nodes, such as those with high volumes of originating transactions. By capturing only a small amount of traffic from these key nodes, a significant portion of their originating transactions can be identified.

\section{COUNTERMEASURES}
\label{sec-fix}

Our solution compromises transaction anonymity and user privacy. We propose two mitigation strategies.

\smallskip
\noindent\textbf{Using Tor or VPN.} The goal of network-layer deanonymization is to identify the IP address of the node originating the transaction. However, if the node uses Tor or a VPN to obscure its real IP address, it can effectively prevent the attacker from successfully deanonymizing the node. Although the attacker may associate the transaction with the VPN provider's IP address, the node's true identity remains concealed.

\smallskip
\noindent\textbf{Upgrading new privacy protection protocols.} Our deanonymization solution is not applicable to the Dandelion++ protocol~\cite{fanti2018dandelion++} and the Erlay protocol~\cite{naumenko2019erlay}. The stem phase in Dandelion++ and the low-fanout flooding phase in Erlay only propagate transactions through peers in outgoing connections, rendering probe-based traffic collection ineffective. However, while Dandelion++ and Erlay provides enhanced anonymity, it also introduces a propagation delay of several seconds.

\section{Related Work}
\label{sec-rw}

\noindent\textbf{Deanonymization in Bitcoin Networks.} Existing studies fall into two typical ways: \textit{transaction layer} and \textit{network layer}. 

\textit{Transaction layer} solutions~\cite{androulaki2013evaluating, meiklejohn2013fistful, moser2022resurrecting, ron2013quantitative, toyoda2017identification, kappos2022peel} focus on deanonymizing transaction addresses by establishing links between anonymous addresses and real-world entities. A common way is to trace the change address. Each transaction generates a single change address, which must return the remaining balance to the sender, providing a potential way to link multiple addresses to the same user.

\textit{Network layer} approaches \cite{kaminsky2011,gao2018,koshy2014analysis,biryukov2014deanonymisation,bojja2017dandelion,fanti2017anonymitypropertiesbitcoinp2p,fanti2018dandelion++,biryukov2019deanonymization} trace transactions by analyzing network traffic and message transmission within the blockchain. By examining the transaction's propagation path, these methods (Table~\ref{tab:deanonymization_methods}) can identify the origin node and trace its associated IP address, linking the transaction back to its source.

\smallskip
\noindent\textbf{Deanonymization in crypto-anonymous networks.}
Ways to deanonymize transactions and nodes in anonymous networks (e.g., those using Tor) are more challenging than in public networks due to the added privacy layers. However, several studies demonstrated vulnerabilities that can be exploited to reveal user identities.

Biryukov et al. \cite{biryukov2015bitcoin} introduced a method allowing a low-resource attacker to control the flow of information between Bitcoin users on Tor. Their approach clusters a user’s transactions and manipulates the relaying of Bitcoin blocks and transactions, enabling the attacker to delay or discard these transmissions. 
Gao et al. \cite{gao2021two} developed a technique to pinpoint the IP addresses of full nodes operating as hidden services on Bitcoin and to identify their originating transactions. Yang et al. \cite{yang2023evicting} exploited Bitcoin's connection eviction mechanism to infer the IP addresses of Bitcoin nodes using the Tor network. Shi et al. \cite{shi2024deanonymizing} introduced the first deanonymization method targeting Monero transactions over Tor.

\begin{table}[!]
\centering
\caption{Detailed \textit{Network layer} Solutions}
\label{tab:deanonymization_methods}
\resizebox{\linewidth}{!}{
\begin{tabular}{c|p{7.5cm}}
\toprule
\multicolumn{1}{c}{\textbf{}} & \multicolumn{1}{c}{\textbf{Key Method and Results}} \\
\midrule

Kaminsky~\cite{kaminsky2011} & \cellcolor{gray!15} First node to forward a transaction is likely the originator, but accuracy is low due to reliance on this single feature. \\

Koshy~\cite{koshy2014analysis} & Linked abnormal forwarding behaviors to Bitcoin addresses, but the limited occurrence of such behaviors reduces effectiveness. \\

Biryukov~\cite{biryukov2014deanonymisation} & \cellcolor{gray!15} Identified originating nodes by tracking outgoing connections, improving on previous methods by using post-NAT IPs. \\

Feng~\cite{gao2018} & Developed a lightweight method using active sniffing to trace transaction paths and link Bitcoin addresses to IPs. \\

Fanti~\cite{fanti2017anonymitypropertiesbitcoinp2p} & \cellcolor{gray!15} Showed that the diffusion mechanism does not sufficiently protect anonymity, and proposed Dandelion. \\

Biryukov~\cite{biryukov2019deanonymization} & Proposed transaction clustering based on diffusion, narrowing source IPs and achieving high precision/recall for Bitcoin/Zcash. \\

\bottomrule
\end{tabular}
}
\end{table}

\smallskip
\noindent\textbf{Deanonymization with machine learning.} Apostolaki et al. \cite{apostolaki2021perimeter} introduced the use of machine learning for transaction deanonymization at the network layer. Their attack leveraged the interaction between the network and application layers, employing a routing attack where an AS-level listener intercepts packets sent by the victim node without detection. They applied an unsupervised learning algorithm to identify transactions, marking the first time such an approach was used in this context. Their experiments demonstrated that an AS-level attacker intercepting 25\% of a victim node's connections could deanonymize transactions originating from that node with 70\% accuracy. However, the study did not provide a quantitative analysis of deanonymization precision. Furthermore, since various abnormal behaviors could be misclassified as noise, unsupervised learning algorithms struggled to differentiate the target behavior from irrelevant anomalies.

In comparison, we are the first to leverage semi-supervised learning in deanonymization, and obtain better performance.

\smallskip
\noindent\textbf{Deanonymization with collaborative analysis.} 
Neudecker et al.~\cite{neudecker2017could} performed a correlation analysis between address clustering and IP address information collected at the network layer to investigate the relationship between address clusters and IP addresses.

In contrast, we leverage transaction clustering results from the transaction layer to enhance the network-layer deanonymization results obtained through NTSSL/NTSSL+.


\section{Conclusion}
\label{sec-conclu}

We present NTSSL and NTSSL+, two deanonymization methods based on semi-supervised learning. NTSSL is a low-cost approach that eliminates the need for labeled data by generating pseudo-labels through unsupervised learning algorithms and oversampling techniques. When 25\% of a target node's connections are intercepted, NTSSL achieves a recall rate of 56.8\% and a precision rate of 50\%. We then establish NTSSL+, a cross-layer collaborative deanonymization scheme that integrates transaction clustering to enhance NTSSL's accuracy. NTSSL+ attains a recall rate of 72.7\% and a precision rate nearing 60\%, outperforming PERIMETER (FC21) even with only 25\% of the target node's connections intercepted.


\bibliographystyle{unsrt}
\bibliography{bib}

\end{document}